# Thermalization of neutrons at ultracold nanoparticles to the energy range of ultracold neutrons

V.V.Nesvizhevsky
Institut Laue-Langevin

**Abstract**
Two hypothesizes concerning the interaction of neutrons with nanoparticles with implications for the physics of ultracold neutrons (UCN) were recently considered in ref. [1]; they were motivated by the experimental observation of small changes in the energy of UCN upon their collision with the surface. The first hypothesis explains the nature of the phenomenon observed during the inelastic coherent scattering of UCN on contact with nanoparticles or nanostructures weakly attached at the surface, in a state of permanent thermal motion. It has received experimental confirmation in ref. [2]. The second hypothesis reverses the problem of neutron interaction with nanoparticles in the following manner. In all experiments with UCN, the trap-wall temperature was much higher than 1 mK, which corresponds to the UCN energy. Therefore, UCN preferentially increased their energy. The surface density of weakly attached nanoparticles was low. If, however, the temperature of the nanoparticles is lower than the temperature of the neutrons and if the density of the nanoparticles is high, the problem of the interaction of neutrons with nanoparticles is inverted. In this case, the neutrons can cool down, under certain conditions, to a temperature of about 1 mK, owing to their scattering on ultracold-heavy-water, deuterium, and oxygen nanoparticles, with the result that the UCN density increases by many orders of magnitude. In the present article we repeat the argument used in ref. [1] and formulate in a general way a research program for verifying the validity of this hypothesis. Both the theoretical and experimental activity around the investigation of this issue is likely to intensify in the near future.

**Introduction**
A series of experiments performed by a number of research groups have brought to light the phenomenon of the quasi-elastic scattering of UCN displaying surprisingly small energy changes in the order of ~$10^{-7}$ eV [3-9]. The studies were inspired by the long-standing problem of the excessively high loss of neutrons from traps, a problem that has been repeatedly reported ever since the first experiment on UCN storage in traps [10]. Numerous studies, such as ref. [11] and many other related publications, have failed to provide a clear explanation, highlighting only the insufficiency of our knowledge of the process. It is important to address this question given the wide-spread application of UCN storage techniques to neutron-based research in fundamental particle physics, including the measurement of the neutron lifetime [12-16], the search for the non-zero neutron electric dipole moment [17-18], the study of the gravitationally bound quantum states of neutrons [19-21], and the search for the non-zero neutron electric charge [22].

Our recent observations of UCN loss from traps were as follows: when reflected from a sample surface or the spectrometer walls, UCN change their energy (increasing it by preference) by about $10^{-7}$ eV with a probability of $10^{-7}$-$10^{-5}$ per collision. We refer to this process as the "small heating" of UCN, or, by analogy with the common evaporation process, as "VUCN formation" (Vaporizing UCN). If the resulting neutron energy is higher than the wall's potential barrier, the neutron can penetrate the wall material and either continue directly across, or else be absorbed or up-scattered. A detailed study of this process [23-26, 1-2] has allowed us to conclude that, for solid surfaces at least, this is due to the inelastic coherent scattering of UCN on nanoparticles or nanostructures weakly attached to the surface in a state of permanent thermal motion. Although investigation is still in progress, we are already able to state that a correct account of this process would be important for the proper interpretation of the many contradicting experiments using UCN storage in traps; it would also indicate a method for studying the dynamics of nanoparticles and nanostructures and, accordingly, their interactions with the surface or with one another; the nanoparticles and nanostructures can be selected by size. In any experiment of this kind, the nanoparticle temperature is equal to the trap temperature $T_{trap}$ in the typical range of $10^1$-$10^3$ K, while the UCN energy corresponds to $T_{UCN}$ ~1 mK. Ultracold neutrons therefore preferentially increase their energy in collisions with such "warm" nanoparticles. The probability of such inelastic UCN scattering on the surface is small, since the surface density of such weakly attached nanoparticles is typically small.

However, the mathematical problem of neutron-nanoparticle interaction can in principle be reversed: the interaction of warm neutrons with ultracold nanoparticles at a temperature of ~1 mK can cool down the neutrons. If the density of weakly attached nanoparticles is high (these nanoparticles not only cover the surface but also fill the volume) and if, as the neutrons cool, the probability of their absorption and β- decay is low, the neutron density will increase. This process can, for the first time, allow equilibrium cooling of neutrons down to the UCN temperature.

In order to produce UCN, nuclear fission in nuclear reactors is used, releasing neutrons of energy ~$10^7$ eV. The energy of neutrons in pulsed sources based on proton accelerators is commensurate with that in reactors. However, the cooling of neutrons by a factor of about $10^8$! is achieved by just a few dozen collisions with nuclei in reactor moderators (hydrogen, deuterium). The energy transfer is very efficient, and the neutron losses during the cooling process are low, since the mass of the moderator nuclei is equal to (or approximates) the neutron mass. However, no further cooling occurs: the lower the neutron energy, the larger the neutron wavelength. When the wavelength becomes commensurate with the distance between the nuclei of the moderator, the neutrons do not "see" individual nuclei any longer – they are just affected by the average optical potential of the medium. The neutron energy becomes lower than the bounding energy of the atoms in the medium. A further cooling of the neutrons due to their interaction with collective degrees of freedom (such as phonons) is less efficient than the moderation of the neutrons due to their collisions with nuclei. However, this does allow the cooling of the neutrons to the energy range of cold neutrons (about $10^{-3}$ eV). That is insufficient however to cool the main portion of the neutrons to the UCN energy region [27-32]. The idea of neutron cooling on ultracold nanoparticles consists in reproducing the principle of neutron cooling in reactor moderators via multiple collisions. However, the scale is different: the sizes are greater by a factor of ~$10^2$; this increases the energy range of application of this mechanism by a few orders of magnitude. The energy and the temperature scales, which correspond to the mechanisms being considered, are shown in

Figure 1. It should be noted that such a UCN source is based on the principle of UCN density accumulation, as in a super-thermal source [29], but not on the use of a UCN flux from a source in the flow-through mode. In conventional sources used to select UCN, thermal equilibrium is not achieved. These sources are much hotter than UCN. Only a very small portion of the neutrons is used – other neutrons are lost. Actually, these are sources of cold or very cold neutrons (VCN), and experimentalists have to select a narrow fraction of a broad energy spectrum. For instance, the most intense flux of UCN available for users is now produced in a liquid-deuterium source placed within the core of the high-flux reactor at the Institut Laue-Langevin (ILL) [28]. It increases the UCN flux by a factor of about $10^2$ in relation to that available otherwise in the reactor in the thermal equilibrium spectrum. Only a fraction of the neutron flux of about $10^{-9}$ is thus actually used. On the other hand, the cooling of neutrons on ultracold nanoparticles could provide for further neutron cooling in a significant energy range, thereby increasing the neutron density.

The new method for producing UCN consists in the equilibrium cooling of VCN – through their many collisions with ultracold nanoparticles made from low-absorbing materials ($D_2O$, $D_2$, $O_2$ etc) – down to the temperature of these nanoparticles of ~1 mK, during the diffusion motion of these neutrons in a macroscopically large body of nanoparticles. The principle of equilibrium cooling allows an increase in the neutron phase-space density, in contrast to the method of selecting a narrow energy range out of a warmer neutron spectrum. The use of nanoparticles provides a sufficiently large cross-section for coherent interaction and an inhomogeneity of the moderator density on a spatial scale of about the neutron wavelength; it also shifts the energy transfer range far below a value of about $10^{-3}$ eV, the characteristic limit for liquid and solid moderators. Many collisions are needed since the nanoparticles' mass is much larger than the neutron mass; the energy transfer to nanoparticles and nanostructures is only moderately efficient. The need for a large number of collisions limits the choice of materials: only low-absorbing materials are appropriate. The nanoparticles' temperature must correspond to the minimal energy to which neutrons can still be cooled using this method. The diffusion motion of neutrons in the body of nanoparticles allows us to minimize the thermalization length and, accordingly, to increase the achievable UCN density. The cooling itself is provided by the interaction of neutrons with individual degrees of freedom of weakly bound or free nanoparticles, as well as by the excitation of collective degrees of freedom in the body of nanoparticles (e.g. vibrations and rotations), and also by the breaking of inter-particle bonds.

**Model of free nanoparticles and estimation of parameters**
(i) Let us estimate the loss of neutrons following their capture in nuclei during their cooling in an infinite volume of free-molecule gas. At low temperatures, all gases become liquid (helium) or solid. Therefore, a consideration of neutron scattering on free molecules at temperatures of about 1 mK is only the first step in analyzing neutron interaction with nanoparticles. From the theory of neutron cooling in reactor moderators, it follows that, for an isotropic angular distribution of scattered neutrons in the c.m. frame, the cooling of neutrons in gases of free atoms (or molecules) with an atomic mass $A$ is efficient if:

$$\frac{\sigma_{coh}}{A\sigma_{abs}} > \ln\left(\frac{V_i}{V_f}\right) \tag{1}$$

where $V_i$ is the initial neutron velocity and $V_f$ is the final neutron velocity. It should be noted that the coherent-scattering cross-section $\sigma_{coh}$ is independent of the neutron velocity, while the absorption cross-section $\sigma_{abs}$ is proportional to the reciprocal neutron velocity:

$$\sigma_{abs}(V_n) \sim \sigma_{abs}(V_0) V_0 / V_n \tag{2}$$

This circumstance limits the minimal velocity $V_{min}$ that can be achieved owing to the cooling of neutrons in a free molecular gas even at zero temperature. On the other hand, losses of neutrons are negligible when the neutron velocity is higher than this critical limit. The condition in (1) constrains the list of candidates to a very few: deuterium, oxygen, probably carbon, or a combination of these atoms. Table 1 below compares different materials for nano-moderators. The cooling of neutrons down to velocities even lower than that presented in the Table is not efficient: fortunately, however, such neutrons have already been sufficiently cooled in order to be trapped.

| Molecule/atom | $V_{min}$, m/s |
|---|---|
| D, D$_2$ | ~0.4 |
| D$_2$O | ~1.0 |
| O, O$_2$ | ~2.4 |
| CO$_2$ | ~10. |
| C | ~16 |
| Be | ~20 |

*Table 1. Estimates of the minimal velocity to which neutrons can be cooled in a gas of free atoms, molecules, or nanoparticles of various materials (the condition $\left[\frac{\sigma_{coh}}{A\sigma_{abs}(V_{min})} = 1\right]$).*

The estimations given in the Table 1 are valid for all interesting practical cases, as they are model-independent. However, the thermalization length $L_{therm}$ and thus the characteristic size of such a moderator would be unrealistically large, being of the order of many meters in the very best of cases.

(ii) Evidently, for an ultracold moderator sized within practical limits, any decrease in the thermalization length increases the density of cooled neutrons. This requires a significant increase in the neutron-scattering cross-section of the moderators; this may be achieved by assembling molecules (or atoms) into nanoparticles. The cross-section of the interaction of neutrons with nanoparticles is proportional to the square of the number of molecules in a nanoparticle. With increasing nanoparticles size $d$, there arise two factors that compensate each other. (A) The number of collisions needed for cooling neutrons is proportional to the nanoparticles' mass $M$. (B) The ratio of the coherent-scattering cross-section to the absorption cross-section is also proportional to the nanoparticles' mass, $\sigma_{coh}^M / \sigma_{abs}^M \sim M$. Therefore, the condition in (1) for efficient cooling is valid until the nanoparticles' size compared to the neutron wavelength $\lambdabar_n$ becomes so great that scattering proves to be anisotropic; that is, if $d < \lambdabar_n$, then:

$$\frac{\sigma_{coh}^M}{M\sigma_{abs}^M} \approx \frac{\sigma_{coh}}{A\sigma_{abs}} \qquad (3)$$

If the neutron velocity is higher, or if the nanoparticles' size exceeds these limits (or if both these conditions are satisfied), the angular distribution of scattered neutrons is directed forward. This change in the angular distribution of scattered neutrons increases the relative importance of absorption. In this case condition (3) becomes invalid, since the coherent-scattering cross-section increases more slowly than in proportion to $N^2$; also, the energy transfer per collision decreases. Therefore, the velocity range for neutrons that can still be cooled is restricted at the top end of the range by a velocity maximum $V_{max}$. Evidently, such order-of-magnitude estimations have to be confirmed by precise quantum-mechanical calculations of the energy (velocity) - dependence of the neutron-nanoparticle interaction. This can be easily done and will be published soon.

(iii) Thus, the neutron velocity range $V_{min} - V_{max}$ in which efficient cooling by collisions with ultracold nanoparticles is possible is restricted at both ends of the range: the minimum velocity $V_{min}$ is restricted by neutron absorption in the nuclei of the nanoparticle material, while the maximum velocity $V_{max}$ is restricted by a decrease both in the interaction cross-section and in the energy transfer. The broader this ranges of acceptable velocities, the greater the resulting increase in the neutron phase-space density. However, one should note that if $V_{max}$ is significantly larger than $V_{min}$, one could obtain high UCN density by repeating the cooling cycle with the same neutrons. The resulting UCN density would depend in this case only on the maximum heat load acceptable for the cryogenic system.

(iv) Nevertheless, the range of acceptable neutron velocities in the model of free nanoparticles is broad. An estimate of $V_{min}$ is independent of the nanoparticles' size, since the condition in (3) will always be valid if $d < \lambdabar_n$. But the neutron wavelength is proportional to the reciprocal neutron velocity and is large at the final stage of cooling, at the low-velocity limit. Thus, it is only the nanoparticle material that determines the value of $V_{min}$, which can be estimated as:

$$\frac{\sigma_{coh}}{A\sigma_{abs}(V_{min})} = 1 \qquad (4)$$

That is, $V_{min}$ can be as low as about 1 m/s (see Table).

A solution to the quantum mechanical problem of neutron interaction with a nanoparticle is beyond the scope of the present study. However, in the simplified model considered in ref. [1], the thermalization length for heavy-water nanoparticles was:

$$L^{D_2O}[cm] \approx \frac{40}{(d[nm])^{3/2}} \qquad (5)$$

and for a reasonable moderator size of ~10 cm and nanoparticle diameter of ~2.5 nm, the maximum allowed initial neutron velocity $V_{max}$ was ~25 m/s, or even ~$10^2$ m/s.

**Realistic moderators**

(i) In real nanoparticle moderators at ultra-low temperatures, nanoparticles are not free. However, the interaction between them can be very weak. For example, if one takes nanoparticles of the required size and material ($D_2O$, $D_2$, $O_2$ etc) and drops them into liquid $^4$He (not absorbing neutrons and just providing heat transfer), they are immediately coated with a thin layer of solidified helium. This layer screens the nanoparticles from one another and reduces the interaction between neighboring nanoparticles [33-35]. Such gels of ultracold nanoparticles provide an unique opportunity for the extremely efficient cooling of neutrons, as they possess all the properties required: weak bonds between the nanoparticles, nanoparticle size close to the optimum, high porosity providing high contrast of the corresponding effective potential for scattering of neutrons, low absorption of neutrons if the nanoparticle material is properly chosen. There remains an open question: Does the inter-nanoparticle interaction in such gels leave sufficient freedom for them? Their independent interaction with neutrons is of crucial importance, since, without it, the effective mass of nanoparticles increases, with the result that the energy transfer decreases dramatically. On the other hand, additional degrees of freedom (vibrations, rotations, and breaking of interparticle bonds) in such gels probably provide an even more efficient cooling of neutrons than collisions, and they should be given special consideration. This is particularly important because such one-step cooling events could allow us to extend the range of acceptable initial neutron velocities to $10^2$-$10^3$ m/s, and thus to approach an ideal moderator with 100% efficiency, cooling all initial neutrons down to the UCN energy range. The information on the collective degrees of freedom and structure of the gels could be obtained in various ways: by direct first-principle modeling of the formation of the gels; using known non-neutron methods for the gel structure; by measurement of cross-sections of inelastic interaction of neutrons in a broad range of initial velocities with small gel samples – for the investigation of the collective degrees of freedom in the gels. All these research paths will be pursued in the near future.

(ii) In real moderators, in contrast to the simplified model of free nanoparticles, it is necessary to take into account neutron-optical effects due to the finite distances between nanoparticles. As soon as the neutron energy becomes sufficiently low (the neutron wavelength becomes sufficiently large), the neutron wavelength simultaneously covers a few nanoparticles; there is therefore an effect of coherent upward scattering of neutrons in the body of nanoparticles. This results in the following: (A) the energy transfer decreases and the cooling process become less efficient; (B) the depth of extraction of such neutrons increases, and the moderator becomes more transparent for such low-energy neutrons. In this case, UCN can spread out to the moderator surface from its total depth, and this simplifies their extraction. In order to get information on this problem, all the methods mentioned in the previous section plus the measurements of the elastic neutron cross-sections need to be employed.

(iii) The technical feasibility of such a moderator needs to be carefully studied. The use of ultra-low temperatures does not allow it to be placed in the vicinity of the reactor core, as the nanoparticles and the cryostat would be heated by neutrons and especially by accompanying gammas. It can be installed at the exit of an optimized VCN or cold neutron guide, although even then heating is a major problem. The particular design of such a moderator depends on results of measurements of elastic and inelastic cross-sections of interaction of neutrons with the gel samples of various compositions and structures. All preliminary investigations of the corresponding elastic and inelastic cross-sections could be carried out using an optical cryostat at the temperature of ~1 K using small gel samples. This would allow the feasibility in principle of such neutron thermalization to be demonstrated and the production of UCN to be optimized. It would define the transport characteristics of neutrons in nano-structured matter and give an idea of their thermalization efficiency. At the second stage of this project, however, a full-scale ultra-low-temperature cryostat and large gel samples would be required in order to study the thermalization process itself.

**Conclusion**

A new concept for producing high UCN density is proposed. This concept is based on the cooling of neutrons on ultracold nanoparticles of heavy water, deuterium, or oxygen in super-fluid helium. Based on the model of free nanoparticles, this method can be applied at initial neutron energies in the range of $10^{-8}$-$10^{-7}$ eV to $10^{-5}$-$10^{-4}$ eV. A precise account of the collective degrees of freedom (vibrations, rotations of chains of nanoparticles, breaking of nanoparticle bonds) would allow us to extend the initial neutron energies to the $10^{-4}$-$10^{-2}$ range. There is no obvious limitation in principle to increase the UCN density in such a way, except for the finite acceptable heat load on the ultra-low-temperature cryostat associated with neutrons. This method differs from traditional methods for UCN production in the high efficiency of employing the initial neutron flux. There is a need for a more detailed theoretical and experimental study of such a cooling process, as well as for reliable estimations of the UCN density gain and the maximum neutron energy, at which such a process is still efficient.

**Aknowledgements**

I am very grateful to all my colleagues who have contributed to or expressed their interest in this project, and in particular to: I.S.Altarev, A.M.Barabanov, S.T.Belyaev, H.G.Börner, V.E.Bunakov, V.B.Efimov, V.P.Gudkov, V.V.Khmelenko, G.V.Kolmakov, D.M.Lee, A.A.Levchenko, A.V.Lokhov, E.V.Lychagin, L.P.Mezhov-Deglin, V.I.Morozov, A.Yu.Muzychka, S.Paul, J.M.Pendlebury, A.K.Petukhov, K.V.Protasov, G.Quemener, D.Rebreyend, A.V.Strelkov, and E.Tomasi-Gustaffson.

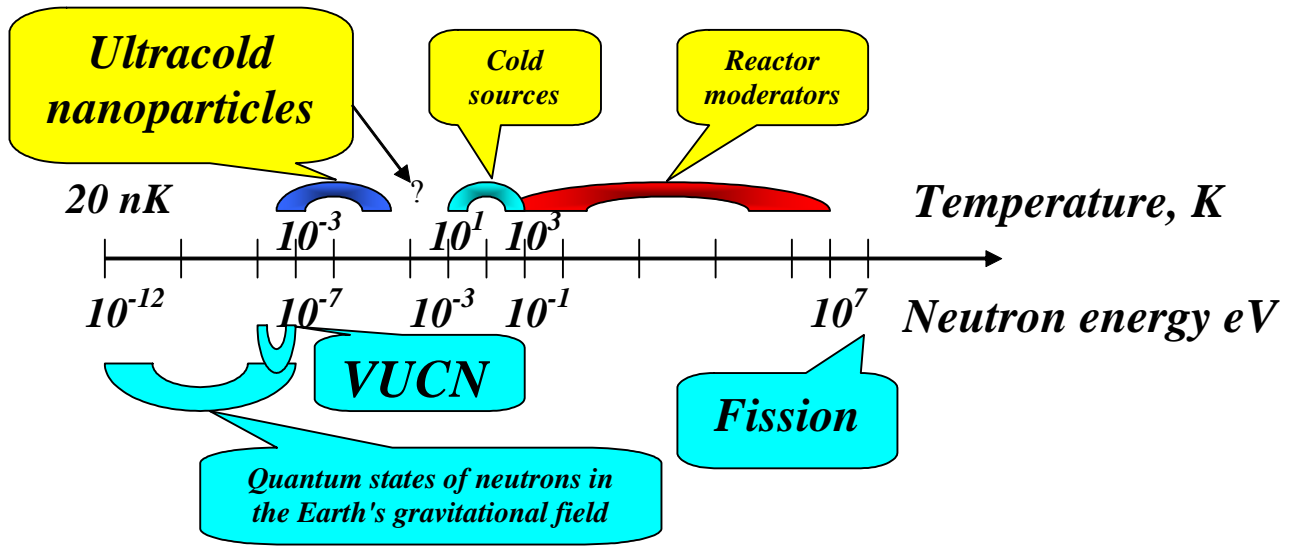

*Figure 1. Neutron energy and temperature ranges that correspond to various moderators, along with a few examples of physical phenomena involving UCN and even slower neutrons.*